\documentclass[a4paper]{article}
\usepackage{graphicx,epsfig,rotating}
\newcommand{\ra}{$\;\rightarrow$}
\begin{document}
 

\begin{titlepage}

\vspace*{15mm}
\begin{flushright}{\large\bf July 7, 1999}\end{flushright}

\vspace*{30mm}

\begin{center}
{\LARGE\bf The $H_{SUSY}$ \ra $~\tau\tau$ \ra $~h^{\pm}+h^{\mp} + X$ channel, its advantages and potential instrumental drawbacks}
\end{center}

\vspace*{15mm}

  \begin{center}
    R.~Kinnunen~$^{a)}$\\
       Helsinki Institute of Physics, Helsinki, Finland\\
\vspace{4mm}
    D.~Denegri~$^{b)}$\\ 
       DAPNIA/SPP, CEN Saclay, France\\
\end{center}

\vspace{20mm}
  \begin{abstract}

We present a first study of the channel $H, A$ \ra $~\tau\tau$ \ra $~h^{\pm}+ h^{\mp} + X$  in CMS at high $m_A$ values where no triggering difficulties are expected with QCD jets. At present the $\tau$ selection is based solely on the presence of a hard isolated track in the "$\tau$" jet, but further refinements based on calorimeter collimation or impact parameter selections are obviously possible. The main irreducible background in these conditions is due to QCD jets with hard fragmentations. A large reduction of this background and improvement in the expected signal to background ratio is provided by $E_t^{miss}$ cuts. The expected high-mass reach in the $m_A$, $tan\beta$ parameter space for $3 \times 10^4pb^{-1}$ is shown. This $H$ \ra $~\tau\tau$ channel provides the highest mass reach and the best mass resolution  when compared to $\tau\tau$ \ra $~l^{\pm}+h^{\mp} + X$
and $\tau\tau$ \ra $~e^{\pm}+\mu^{\mp} + X$ final states. To the extent that with further calorimetric and impact parameter based selection criteria the QCD background can be kept under control, i.e. below the irreducible $Z,\gamma*$ \ra $~\tau\tau$ background, we should strive to have a first level trigger allowing to explore the mass range down to $\sim$150 - 200 GeV.
\end{abstract}

\vspace{20mm}
\hspace*{-8mm} \hrulefill \hspace{60mm} \hfill \\
$^{a)}$  Email: ritva.kinnunen@cern.ch\\
$^{b)}$ Email: daniel.denegri@cern.ch  

\end{titlepage}
\pagenumbering{arabic}
\setcounter{page}{1}


\section{Introduction}

We present here the first investigation of the purely hadronic $\tau$ decay modes in our systematic investigation of $h,H,A(H_{SUSY}$) \ra $~\tau\tau$ channels. The one lepton plus one charged hadron \cite{tausel},\cite{tauhlep} and the one electron plus one muon \cite{tauemu} final states have been shown to cover a significant region of the $m_A$, $tan\beta$ parameter space at high $tan\beta$ values ($tan\beta >$ 10 and $m_A>$ 150 GeV). These channels have been studied in both the inclusive $H_{SUSY}$ production mode, and in $b\overline{b}H_{SUSY}$ in association with b-jets. For Higgs masses from 200 to 800 GeV the fraction of events produced in association with b-jets varies from 75\% to 80\% for $tan\beta$ = 10 and from 75\% to 96\% for $tan\beta$ = 45. The associated production channels provide much better signal to background ratios, but statistics is much reduced (factor $\sim$20) and are very sensitive to the tracker performance through the b-tagging procedure, the associated b-jets tending to be rather soft and uniformly distributed between barrel and endcap pixels \cite{btag}. Studies of these channels are continuing, implementing detailed pattern recognition and track finding in the CMS tracker \cite{tdr:tracker}.

For these $A, H$ \ra $~\tau\tau$ \ra $~h^+$ + $h^-$ final states with two isolated hard hadrons (plus $\pi^0$'s) we start with high masses $m_A >$ 300 GeV, as for the low mass range ($\sim$ 150 GeV) these purely hadronic final states competing with QCD jets obviously run into difficulties firstly of triggering at an acceptable rate, and second, in the off-line analysis on problems of large background from QCD jets followed by hard fragmentation. These channels can however provide a spectacular signature for a sufficiently heavy Higgs and should obviously extend significantly the upper $H_{SUSY}$ mass reach due to favourable branching ratios. The branching ratio of $\tau$ to single hadron plus any number of neutrals is about 50\%. The combined branching ratio for $A, H$ \ra $~\tau\tau$ \ra $~h^+ + h^- + X$ is 2.8\%  compared to about 3.7\% for  $A, H$ \ra $~\tau\tau$ \~ra $l^{\pm} + h^{\mp} + X$ assuming BR($A, H$ \ra $~\tau\tau$) = 11\% being roughly valid for the full $m_A$ and $tan\beta$ ($>$10) range studied here. The fractional momentum taken away by the neutrinos is significantly smaller in the $h^+$ + $h^-$ final state compared to $l^{\pm} + h^{\mp}$ and especially $e^{\pm} + \mu^{\mp}$ final states, thus an improved $H_{SUSY}$ \ra $~\tau\tau$ mass resolution can be expected, and improved signal/background if QCD background can be reduced below the irreducible $\tau\tau$ background.  

The main reducible background sources are the QCD jets and $W+jet$ events with $W$ \ra $~\tau\nu$, and the irreducible ones are $Z, \gamma^*$ \ra $~\tau\tau$ and $t\overline{t}$ with $W$ \ra $~\tau\nu$. The rate for QCD jets is very large; about $3 \times 10^{11}$ two-jet events with $E_t^{jet} >$ 60 GeV are expected for $3 \times 10^4pb^{-1}$. Therefore a rejection factor of at least $10^4$ per jet is needed for a useful signal (further rejection can be provided by $E_t^{miss}$ cuts). In this study the $\tau$ identification is based solely on the presence of a single hard isolated charged hadron in the jet using tracker information. Indeed a large rejection factor is obtained from the tracker alone, a result expected to be confirmed with the detailed simulation and track reconstruction currently under way. The use of calorimeter information exploiting the collimation of a $\tau$ jet to improve $\tau$ identification is also presently in progress, with GEANT calorimeter simulation \cite{tdr:ecal}\cite{tdr:hcal}. We are also investigating the possibility to use the impact parameter measurement of the hard hadron from $\tau$ to further reduce the QCD background. All these improvements will thus further reinforce and extend the reach of the $H_{SUSY}$ \ra $~\tau\tau$ channel, the only channel which until now gives us access to the theoretically favoured $H_{SUSY}>$ 500 GeV mass range. However, this channel should also play an important (decisive?) role in the intermediate mass range 150 GeV $< m_H <$ 300 GeV, which is the most critical one from the point of view of the MSSM $tan\beta$, $m_A$ parameter space coverage, provided we can trigger on it efficiently, which requires a dedicated study and is not discussed here. 
 
\section{Event simulation}

Events are generated with PYTHIA \cite{PYTHIA}. Masses and couplings of SUSY Higgses are calculated using two-loop/RGE-improved radiative corrections \cite{gunion}. The branching ratios and cross sections are normalized using the HDECAY program \cite{HDECAY}. No stop mixing is included and the SUSY particles are first assumed to be heavy enough not to contribute to the SUSY Higgs decays. The decays to neutralinos and charginos can however reduce strongly the  $A, H$ \ra $~\tau\tau$ branching ratio at low $tan\beta$, and by a factor of two even at $tan\beta$ = 30 \cite{HDECAY}. For the high $m_A$ and $tan\beta$ region investigated here the effects of sparticle decay modes are nevertheless relatively modest, as mentioned later on. In the case of stop mixing, if the light stop $\tilde{t_1}$ is lighter than or comparable in mass to the top quark, the squark loop effects are expected to suppress significantly the $gg$ \ra $~h, A, H$ production processes \cite{djouadi}. For the large $m_A$ and $tan\beta$ values discussed here this would mean at most a 20\% reduction in the production rate as we are dominated by $gg$ \ra $~b\overline{b}H_{SUSY}$ tree-diagram  production which is not affected by $\tilde{t}_1$ - $t$ interference effects. For the QCD background, PYTHIA two-jet events with initial and final state QCD radiation are generated. A cut $p_t^{q,g} >$ 50 GeV is applied for the hard process at the generation level. The default Lund fragmentation is used. 

The CMS detector response is simulated and the jets and the missing transverse energy are reconstructed with the fast simulation package CMSJET \cite{cmsjet}. The loss of reconstruction efficiency for the hard track due to secondary interactions in the tracker material ($\sim$ 20\% of $\lambda_{int}$), as well as the degradation of the track isolation due to conversions of accompanying $\pi^0$ \ra $~\gamma\gamma$ in tracker, is taken into account only is average, which is sufficient at this stage.

\vspace{ 5mm}
\section{Selection of events}
\vspace{ 3mm}
\subsection{$p_t^h$ threshold and track isolation}
\vspace{ 3mm}

Hard $\tau$ jets are expected in the $A, H$ \ra $~\tau\tau$ \ra $~h^+ + h^- + X$ events at large  $m_A$ as can be seen from Fig. 1 which shows the $E_t$ distribution of hadronic $\tau$ jets for $m_A$ = 300, 500 and 800 GeV and the $E_t$ distribution of the two hardest jets in QCD jet events (Fig.1c).
Events are required to have at least two calorimetric jets with $E_t >$ 60 GeV within $|\eta| <$ 2.5.
The $\tau$ jet canditates are chosen to be the $\tau$ jets for the signal and for the backgrounds with real $\tau$'s, i.e. from $Z,\gamma^*$ \ra $~\tau\tau$ and $t\overline{t}$ events. For the QCD background the $\tau$ jet canditates are taken to be the jets with highest $E_t$. Events are assumed to be triggered with a two-jet trigger with full trigger efficiency for $E_t^{jet} >$ 60 GeV at this stage. For $m_A >$ 300 GeV and a threshold $E_t^{jet} >$ 60 GeV the trigger efficiency is high and the trigger rate acceptable at $L \sim 10^{33}cm^{-2}s^{-1}$ \cite{trigger}. For high luminosity running, and in particular for the lower mass range 150 GeV $< m_A <$ 300 GeV, a better understunding of possible $1^{st}$ level triggers and of trigger efficiency versus $E_t$ is needed and then implemented in this type of study.

A jet is defined to be a "$\tau$" jet if it contains one isolated hard hadron within $\Delta R <$ 0.1 from the calorimeter jet axis. A powerfull tracker isolation can be implemented in CMS thanks to efficient track finding down to low $p_t$ values (0.9 GeV) \cite{tdr:tracker}. Here we require that there in no other track with $p_t >$ 1 GeV than the hard track in a larger cone of $\Delta R <$ 0.4 around the calorimeter jet axis. This isolation criterion is adequate for running at the luminosity of about $10^{33}cm^{-2}s^{-1}$, i.e. without significant event pile-up. Figure 2 shows the transverse momentum for the isolated single track in a jet with $E_t >$ 60 GeV and $|\eta|<$ 2.5 for signal events at $m_A$ = 300, 500 and 800 GeV, for the $Z,\gamma^*$ \ra $~\tau\tau$, $t\overline{t}$ events with $W$ \ra $\tau\nu$ and for QCD jet events. The  $\tau$ selection efficiencies per jet, due to both track isolation and the hard track $p_t$ threshold cut are shown in Table 1 as a function of the $p_t$ for the signal and backgrounds. As can be seen from the table, a rejection factor against the QCD jets varying from 700 to about 1500 is obtained from tracker momenta alone. In the following, $p_t^h >$ 40 GeV is chosen, with an efficiency of 50\% for the signal at $m_A$ = 500 GeV and 22\% and 32\% for the $Z, \gamma^*$ \ra $~\tau\tau$ background in the low and high mass range, respectively. The rejection factor against QCD jets is then about 1500.

The rejection factor against the QCD jets is obviously sensitive to the fragmentation function used to simulate the jets. On this point a more detailed study is needed to understand the hard fragmentation effects in jets. The rejection factor shown in Table 1 is an average factor for $E_t^{jet} >$ 60 GeV.  As the simulation of even one QCD jet event with two jets with $E_t >$ 60 GeV both containing one isolated hard hadron with $p_t >$ 40 GeV is not possible, we have evaluated the rejection factor as a function of $E_t^{jet}$ generating large numbers of QCD jets in several $E_t$ bins. This is shown in Fig. 3 and is used in the simulation and evaluation of the QCD background assuming also factorisation of the two jet fragmentations. For a QCD jet with $E_t$
values between 300 and 350 GeV, for instance, the probability to fluctuate into a "$\tau$" jet with one isolated charged hadron with $p_t >$ 40 GeV is about $2\times 10^{-4}$. 

The two hard tracks from $\tau^+$ and $\tau^-$ in the signal events have an opposite sign while no such strong charge correlation is expected for the QCD jet events. As the charge assignement in CMS is almost 100\% for tracks with $p_t <$ 1 TeV \cite{tdr:tracker}, about half of the QCD background can be removed by requiring an opposite charge for the two isolated hadrons.

\begin{table}
\centering
\caption{$\tau$ selection efficiency per jet using just tracker momentum measurement for $A,H$ \ra\ $~\tau \tau$ with $\tau$ \ra $~h^{\pm}+X$ at $m_A$ = 300, 500 and 800 GeV, for $Z, \gamma^*$ \ra\ $~\tau \tau$ in the low and high mass range and for QCD jet background, with $E_t^{jet} >$ 60 GeV and $|\eta^{jet}| <$ 2.4. The hard track is required to be within $\Delta R$ = 0.1 from the calorimeter jet axis and isolated in the tracker with respect to tracks with a $p_t$ threshold of 1 GeV in a cone of $\Delta R$ = 0.4 around the calorimeter jet axis.}

\vskip 0.1 in
\begin{tabular}{|c|c|c|c|}
\hline
 Process  & $p_t^h >$ 20 GeV & $p_t^h >$ 30 GeV & $p_t^h >$ 40 GeV \\
\hline
\hline
$A,H$ \ra\ $~\tau \tau$, $m_A$=300 GeV & 38.1 \% & 32.5\% & 27.3\% \\
$A,H$ \ra\ $~\tau \tau$, $m_A$=500 GeV & 61.7 \% & 55.7\% & 50.0\% \\
$A,H$ \ra\ $~\tau \tau$, $m_A$=800 GeV & 65.7 \% & 61.2\% & 56.9\% \\
$Z, \gamma^*$ \ra\ $~\tau \tau$, 130 GeV$<m_{\tau\tau}<$300 GeV & 30.9\% & 25.0\% & 22.2\% \\
$Z, \gamma^*$ \ra\ $~\tau \tau$, $m_{\tau\tau}>$300 GeV & 41.7\% & 36.7\% & 32.1\% \\
$t\overline{t}$, $W$ \ra $~\tau\nu$ & 25.0\% & 20.5\% & 16.6\% \\
QCD jets, $p_t^{q,g} >$ 60 GeV & 1.4$\times10^{-3}$ & 9.4$\times10^{-4}$ & 6.7$\times10^{-4}$ \\
\hline
\end{tabular}
\label{table1}
\vspace{ 3mm}
\end{table}

\vspace{ 3mm}
\subsection{$E_t^{miss}$ cuts}
\vspace{ 3mm}

Due to the large Higgs masses considered here the missing transverse energy $E_t^{miss}$ from the neutrinos is expected to be significant, although the two $\tau$'s tend to be in back-to-back configuration resulting in a partial compensation. To perform $H$ \ra $~\tau\tau$ mass reconstruction non-back-to-back $\tau\tau$ pairs are required. 
Figure ~\ref{fig:4} shows the expected $E_t^{miss}$ from the simulation of the CMS calorimeter response with the CMSJET program for a signal with $m_A$ = 300, 500 and 800 GeV and for the QCD jet background, in a data taking regime without any event pile-up. After requiring two $\tau$ jets, a cut $E_t^{miss} > $ 40 GeV reduces the QCD background by a factor of about 100, while the 
efficiency for the signal is 59\% at $m_A$ = 500 GeV and 33\% at $m_A$ = 300 GeV. For the largest masses the cut can 
be increased to $E_t^{miss} > $ 60 GeV with a rejection factor against QCD of 180 and an efficiency of 58\% for the signal. These numbers are evidently rather sensitive to CMSJET modelling of detector response and the no pile-up hypothesis, but also to the exact $H_{SUSY}$ production mechamism etc. The presence of the forward calorimetry (VFCAL) is particularly important for the $E_t^{miss}$ measurement in this channel as illustrated in Fig. ~\ref{fig:4}b, showing the $E_t^{miss}$ distributions for QCD events with two jets with $E_t>$ 60 GeV for the full $\eta$ range and for the central calorimeters only ($|\eta|<$ 3). The $E_t^{miss}$ distributions are also significantly modified in the presence of event pile-up in the $E_t^{miss}\sim$ 10 -30 GeV range, which is discussed in a separate note \cite{higlum}, but a cut at $E_t^{miss}>$ 40 GeV is rather safe.
  
\vspace{ 3mm}
\subsection{Relative azimuthal distributions}
\vspace{ 3mm}

The two $\tau$ jets in the signal events are predominantly in the back-to-back configuration,  especially at high masses, as can be seen from Fig. 5 showing the $\Delta\phi$ angle in the transverse plane between the two $\tau$ jets in the signal at $m_A$ = 300, 500 and 800 GeV (Figs. 4 a, b, c).  The $\Delta\phi$ angle for the $Z, \gamma^*$ \ra $~\tau\tau$,  $t\overline{t}$  and QCD backgrounds is shown in Figs. 4 d, e and f. Only the $t\overline{t}$ background component is significantly less back-to-back correlated.

\begin{table}
\centering
\caption{Cross section times branching ratio and number of events for $A,H$ \ra\ $~\tau \tau$ at $m_A$ = 300 and 500 GeV and for the background processes for $3 \times 10^4pb^{-1}$. Shown are the number of events after selection ot two jets with $E_t>$ 60 GeV and $\tau$ identification in the tracker, $\Delta\phi$ cut, $E_t^{miss}$ cut and the reconstruction of Higgs mass. The two hadrons in the $\tau$ jets are required to be of opposite sign. A reconstruction efficiency of 85\% per jet is assumed for $\tau$ jet reconstruction, consistent with CMSIM evaluation.}

\vskip 0.1 in
\begin{tabular}{|c|c|c|c|c|c|c}
\hline
 Process  & $\sigma$ * BR  &  "$\tau$" jets  & $E_t^{miss}>$40 GeV & $\Delta\phi<175^0$ & mass rec.\\
\hline
\hline
$A,H$ \ra\ $~\tau \tau$, $m_A$=300 GeV& 215 fb& 574 & 188 & 133 & 70 \\
 $tan\beta$ = 15 &  &  &  & & \\
$A,H$ \ra\ $~\tau \tau$, $m_A$=500 GeV& 44.4 fb& 329 & 195 & 108 & 66 \\
 $tan\beta$ = 20 &  &  &  &  & \\
$Z,\gamma^*$ \ra\ $~\tau \tau$, $m_{\tau\tau}>$130 GeV & 3.08 pb & 818 & 258 & 124 & 80\\
QCD jets, $p_t^{q,g} >$ 60 GeV & 11.0 $\mu$b & 6579 & 53 & 46  & 8  \\
$t\overline{t}$, $W$ \ra $~\tau\nu$ & 1.74 pb & 73 & 52 & 46 & 20 \\
$W+jet$, $W$ \ra $~\tau\nu$  & 949 pb & 410 & 240 & 212 & 20 \\
\hline
\hline
\end{tabular}
\label{table2}
\vspace{ 3mm}
\end{table}

\begin{table}
\centering
\caption{The same as in Table 2 but for $A,H$ \ra\ $~\tau \tau$ at $m_A$ = 800 GeV and $tan\beta$ = 45 with $E_t^{jet} >$ 100 GeV and $E_t^{miss}>$ 60 GeV.}

\vskip 0.1 in
\begin{tabular}{|c|c|c|c|c|c|c}
\hline
 Process  & $\sigma$ * BR  &  "$\tau$" jets   & $E_t^{miss}>$60 GeV & $\Delta\phi<175^0$ & mass rec.\\
\hline
\hline
$A,H$ \ra\ $~\tau \tau$, $m_A$=800 GeV& 32.8 fb& 213 & 124 & 57  & 37 \\
 $tan\beta$ = 45 &  &  &  &  & \\
$Z, \gamma^*$ \ra\ $~\tau \tau$, $m_{\tau\tau}>$130 GeV & 3.08 pb & 286 & 85 & 42 & 30 \\
QCD jets, $p_t^{q,g} >$ 60 GeV & 11.0 $\mu$b & 1158 & 5  & 5  &  1 \\
$t\overline{t}$, $W$ \ra $~\tau\nu$ & 2.90 pb & 22 & 13  & 12  & 7 \\
$W+jet$, $W$ \ra $~\tau\nu$  & 949 pb & 70 & 54 & 48 & 6 \\
\hline
\hline
\end{tabular}
\label{table3}
\vspace{ 3mm}
\end{table}

\begin{table}
\centering
\caption{Number of events after selection within a mass window cut and statistical significance for $A,H$ \ra\ $~\tau \tau$ at $m_A$ = 300, 500 and 800  GeV and for the background processes for $3 \times 10^4pb^{-1}$. In all cases $tan\beta$ is chosen to be near the limit of observability.}
\vskip 0.1 in
\begin{tabular}{|c|c|c|c|}
\hline
    & $m_A$=300 GeV, $tan\beta$ = 20 & $m_A$=500 GeV, $tan\beta$ = 25 & $m_A$=800 GeV, $tan\beta$ = 45\\
   &  230 GeV$<m_{\tau\tau}<$350 GeV & 400 GeV$<m_{\tau\tau}<$700 GeV & 600 GeV$<m_{\tau\tau}<$ 1 TeV\\
\hline
\hline
Signal & 67 & 58 & 33  \\
$Z, \gamma^*$ \ra\ $~\tau \tau$ & 39  & 49 & 16 \\
QCD jets & 4  & 4  &  0.1  \\
$t\overline{t}$ & 8 & 10  & 4 \\
$W+jet$  & 5 & 9 & 2 \\
\hline
\hline
S / B  & 1.1 & 0.7 & 1.4 \\
$S/\sqrt{S+B}$ & 5.9 & 5.0 & 4.4  \\
$S/\sqrt{B}$ & 8.6 & 6.6 & 6.7  \\
\hline
\hline
\end{tabular}
\label{table4}
\vspace{ 3mm}
\end{table}

\vspace{ 3mm}
\subsection{Higgs mass reconstruction}
\vspace{ 3mm}

As well known, despite at least two escaping neutrinos, the Higgs mass can be approximately reconstructed from the two  $\tau$ jets measured in calorimetry and from the $E_t^{miss}$, when the two $\tau$'s are not exactly back-to-back, by
projecting the $E_t^{miss}$ vector on the directions of the reconstructed $\tau$ jets. This has been shown in connection with 
the $A,H$ \ra\ $~\tau \tau$ \ra $~l^{\pm} + h^{\mp}$ \cite{tauhlep} \cite{hmass} and $A,H$ \ra\ $~\tau \tau$ \ra $~e + \mu$ \cite{tauemu} channels. Figure 6a shows for example the reconstructed Higgs mass for $m_A$ = 500 GeV ($tan\beta$ = 20) with the cuts discussed above and without pile-up.  An upper limit of $\Delta \phi(\tau-jet_1,\tau-jet_2) <$ 175$^o$ is applied.  A gaussian fit to the mass distribution for A and H (which are almost mass degenarate) yields 58 GeV for the resolution ($\sigma$). The mass resolution can be significantly improved by reducing the upper limit in $\Delta \phi$ between the two $\tau$ jets i.e. making them less back-to-back as is shown in Fig. 6b for the same mass. For $\Delta \phi <$ 160$^o$ compared to $\Delta \phi <$ 175$^o$ the resolution improves from 58 to 42 GeV and the tail is reduced. This would however lead to a 55\% loss of signal statistics and is not used here but shows the potential gain in resolution in conditions where statistics would not be the limiting factor. Figures 6c and 6d show the reconstructed Higgs masses for $m_A$ = 300 GeV ($tan\beta$ = 15) and for $m_A$ = 800 GeV ($tan\beta$ = 45), all for $\Delta \phi <$ 175$^o$. The fitted mass resolutions are 33 GeV and 92 GeV, respectively. For $m_A$ = 800 GeV and $tan\beta$ = 45 the natural Higgs width is 43 GeV thus instrumental effects still dominate observed signal width.

Figure 7 summarizes our results on the relative resolution ($\sigma^{gauss}$/$m_H$) for the reconstructed Higgs mass in the $e^{\pm} + \mu^{\mp}$ \cite{tauemu}, $l^{\pm} + h^{\mp}$ \cite{tauhlep} and $h^+ + h^-$ channels. The resolution is best in the $h^+ + h^-$ channel ($\sim$ 10\%) and the worst in the $e^{\pm} + \mu^{\mp}$ channel ($\sim$ 25\%). More exactly, for $m_A$ = 500 GeV and $tan\beta$ = 20, for instance, the mass resolution is $\simeq$11\% for $h^+ + h^-$, $\simeq$16\% for $l^{\pm} + h^{\mp}$ and $\simeq$23\% for $e^{\pm} + \mu^{\mp}$ with always the same upper limit in $\Delta \phi$ . This is due to the fact that the mass resolution is dominated by the precision of the $E_t^{miss}$ measurement, and the resolution is better in the channels with a smaller fraction of energy carried away by neutrinos, i.e. in the case of two hadronic $\tau$ decays. This significantly better mass resolution and thus possibly best signal/background ratio (provided the QCD background is reducible below irreducible $\tau\tau$ background) - not fully appreciated before - also speaks in favour of these double hadronic final states. 

\vspace{ 3mm}
\section{Results}
\vspace{ 3mm}
\subsection{Mass spectra, effects of $E_t^{miss}$ cuts}
\vspace{ 3mm}
  
Figure 8a shows the distribution of the reconstructed Higgs mass for $A,H$ \ra $~\tau\tau$ at $m_A$ = 500 GeV for $tan\beta$ = 20 above the total background before any explicit $E_t^{miss}$ cut is applied at this stage. The QCD background dominates and the signal peak is not even visible over the background distribution. Figure 8b shows the same distribution with $E_t^{miss} >$ 40 GeV. The importance of the $E_t^{miss}$ selection and thus of detector hermeticity in general and the VFCAL in this search is evident. The remaining background is due to $Z, \gamma^*$ \ra $~\tau\tau$, $t\overline{t}$ and $W$+jet events. The potential backgrounds from $b\overline{b}$ events and from WW production with $W$ \ra $~\tau\nu$ are found to be negligible. The statistical fluctuations correspond to the expected statistics for $3 \times 10^4pb^{-1}$. The mass distributions before and after the $E_t^{miss}$ cut for a lighter Higgs with $m_A$ = 300 GeV and $tan\beta$ = 15 is shown in Figs.~\ref{fig:9}a and 9b, and for a very heavy Higgs with $m_A$ = 800 GeV and $tan\beta$ = 45 are shown in Figs.~\ref{fig:9}c and 9d. In all the figures 8a,8b and 9 we are showing $A, H$ \ra $~\tau\tau$ signals close to the observability limit at 3$\times$10$^4$ pb$^{-1}$ (see table 4 for exact signal significancies). Further away from the observation limit signals are very spectacular, as visible in fig. 8c for $m_A$ = 500 GeV and $tan\beta$ = 30 and in fig. 8d for $tan\beta$ = 40.

\vspace{ 3mm}
\subsection{Event rates and signal significance}
\vspace{ 3mm}
 
Table 2 gives the summary of the cross section times the branching ratio and the expected numbers of events for $3 \times 10^4pb^{-1}$ for the signal at $m_A$ = 300, 500 GeV and for the background prosesses, after selection of two $\tau$ jets with $E_t>$60 GeV, with the cuts $E_t^{miss}>$ 40 GeV and  $\Delta \phi<$ 175$^0$ and after  Higgs mass reconstruction. Table 3 shows the same for $m_A$ = 800 GeV with harder cuts $E_t^{jet}>$ 100 GeV and $E_t^{miss}>$ 60 GeV. Table 4 summarizes the the expected number of events and the statistical significance for $3 \times 10^4pb^{-1}$ for the signal at $m_A$ = 300, 500 and 800 GeV and for the background prosesses after all the kinematical cuts and applying a window in the reconstructed Higgs mass. An overall reconstruction efficiency of 85\% per $\tau$ jet is assumed. This number is confirmed by reconstructing a sample of signal events with the CMSIM package. No possible $1^{st}$ level trigger losses have been included, and they may be not negligible in the $m_A$ = 300 GeV  case. For the high mass range investigated here and after $E_t^{miss}$ cuts, the QCD jet background is well below the irreducible $\tau\tau$ background. This suggests that extending this channel to lower Higgs masses is very promising.

\vspace{ 3mm}
\section{Conclusions and future prospects}
\vspace{ 3mm}

We have investigated the possibility to look for the neutral SUSY Higgses A and H in $A, H$ \ra $~\tau\tau$ \ra $~h^{\pm}$ + $h^{\mp}$ + X with hadronic $\tau$ decay modes. Despite the simplifying assumptions made concerning both production dynamics (in particular for QCD background) and approximations made in the fast simulation of detector response, the discovery potential for large $A,H$ masses in this channel is clear. Of course, more detailed studies are needed to confirm and consolidate these preliminary results. Let us remind that the $A,H$ \ra $~\tau\tau$ channels are the only ones up to now giving access to the $m_{A,H} \sim$ 0.5 - 1 TeV mass range which is considered as the most plausible one for the $A$ and $H$. The $A,H$ \ra $~b\overline{b}$ modes may allow it too, but this is even less well understood at present than the $A,H$ \ra $~\tau\tau$ channels in CMS. 

To summarize, we require the presence of two calorimetric jets of $E_t >$ 60 - 100 GeV in $|\eta| <$ 2.4 for trigger. These thresholds are about adequate for running at 10$^{33}$ cm$^{-2}s^{-1}$ but are too low for 10$^{34}$ cm$^{-2}s^{-1}$, which requires further study. 
The $\tau$ selection is based here largely on the tracker, requiring a hard isolated single track with $p_t >$ 40 GeV in the jet. This already provides a large rejection factor against QCD jets, allowing a major suppression of this background. The QCD background can be further reduced by a cut on the missing transverse energy. A rejection factor of about 100 is obtained with a cut  $E_t^{miss} >$ 40 - 60 GeV thus suppressing the QCD background much below the irreducible $Z, \gamma^*$ \ra $~\tau\tau$ background. The effectiveness of these  $E_t^{miss}$ cuts at rather modest $E_t^{miss}$ values has however still to be checked with full CMSIM simulations for full reliability. It has also to be evaluated in case of running at 10$^{34}$ cm$^{-2}s^{-1}$ i.e. in presence of pile-up. Figure 10 shows the 5$\sigma$ significance discovery contours for SUSY Higgses as a function of $m_A$ and $tan\beta$ for 3$\times$10$^4$ pb$^{-1}$ (without pile-up) assuming no stop mixing. The discovery range for the $A, H$ \ra $~\tau\tau$ \ra $~h^{\pm}$ + $h^{\mp}$ + X channel extends down to $tan\beta \sim$ 15 at $m_A$ = 300 GeV and down to $tan\beta \sim$20 at $m_A$ = 500 GeV. For very heavy Higgs at $m_A$ = 800 GeV, $tan\beta$ values of $\sim$ 45 can be probed. Even for these boundary $tan\beta$ values the mass peak may be recognized above the background distribution (figs. 8 and 9); away from the boundary the mass peak can be rather spectacular. These $\tau\tau$ \ra $~h^{\pm}$ + $h^{\mp}$ modes provide the best mass resolution at same $m_H$ and $tan\beta$ when compared to other $\tau\tau$ final states (figs. 8c and 8d).

The 5$\sigma$ discovery limit of Fig. 10 is a preliminary result, optimized for  high masses, and based on a fast simulation of the CMS tracker and calorimeter and on a $\tau$ selection, exploiting only momentum measurements in the tracker. An improvement in the QCD background rejection in the low mass range ($m_H <$ 300 GeV) can be expected by including the calorimeter $\tau$ selection developed in \cite{tausel}  exploiting the collimation of the hadronic $\tau$ signal in the ECAL and the $\tau$ isolation in the ECAL+HCAL. The possibility to use impact parameter $\tau$-tagging for the two high $p_t$ tracks to reduce the QCD background is in progress applying a full track finding procedure. The $\tau$-tagging is not as easy as b-tagging, however the impact parameter selection is here to be applied on fast tracks where multiple scattering is minimal and where the best impact parameter resolution is expected, with an asymptotic $\sigma_{ip} \sim$ 20 $\mu m$ \cite{tdr:tracker} whilst $<c\tau > \sim$ 87 $\mu m$ for $\tau$. These improved $\tau$ selection criteria i.e. QCD jet suppression methods are surely necessary to extend the low-mass reach of this channel towards $m_H \sim$ 150 GeV, but can also be regarded as alternatives or complements to the $E_t^{miss}$ selections - if it turned out that these were not as effective in the final detector as expected here.
 
Exploiting $b\overline{b}H_{SUSY}$ associated production channels with b-tagging opens still other possibilities, as it is the way to reduce efficiently the $Z, \gamma^*$ \ra $~\tau\tau$ background. B-tagging will further reduce the QCD background as well. To extend the Higgs search in this channel to high luminosity running, in particular the b-tagging \cite{tdr:tracker} in associated production channels, requires a separate study at high luminosity. 

Let us remind again that the $A, H$ \ra $~\tau\tau$ modes are the only ones (until now) which give access to the most difficult region of the $m_A$, $tan\beta$ plot (150 GeV $< m_A <$ 300 GeV, $tan\beta <$ 10). It is clear that the low-mass, low-$tan\beta$ reach in this channel ultimately will be trigger limited. Every effort should nonetheless be made so that the $m_A <$ 300 GeV region  could be explored at $L >$ 10$^{33}$ cm$^{-2}s^{-1}$ with this $A, H$ \ra $~\tau\tau$ \ra $~h^{\pm}$ + $h^{\mp}$ + X channel and the possible trigger rate and $E_t^{jet}$ threshold limitations properly assessed.

\begin{figure}[htbp]
\centering
\resizebox{160mm}{210mm}
{\includegraphics{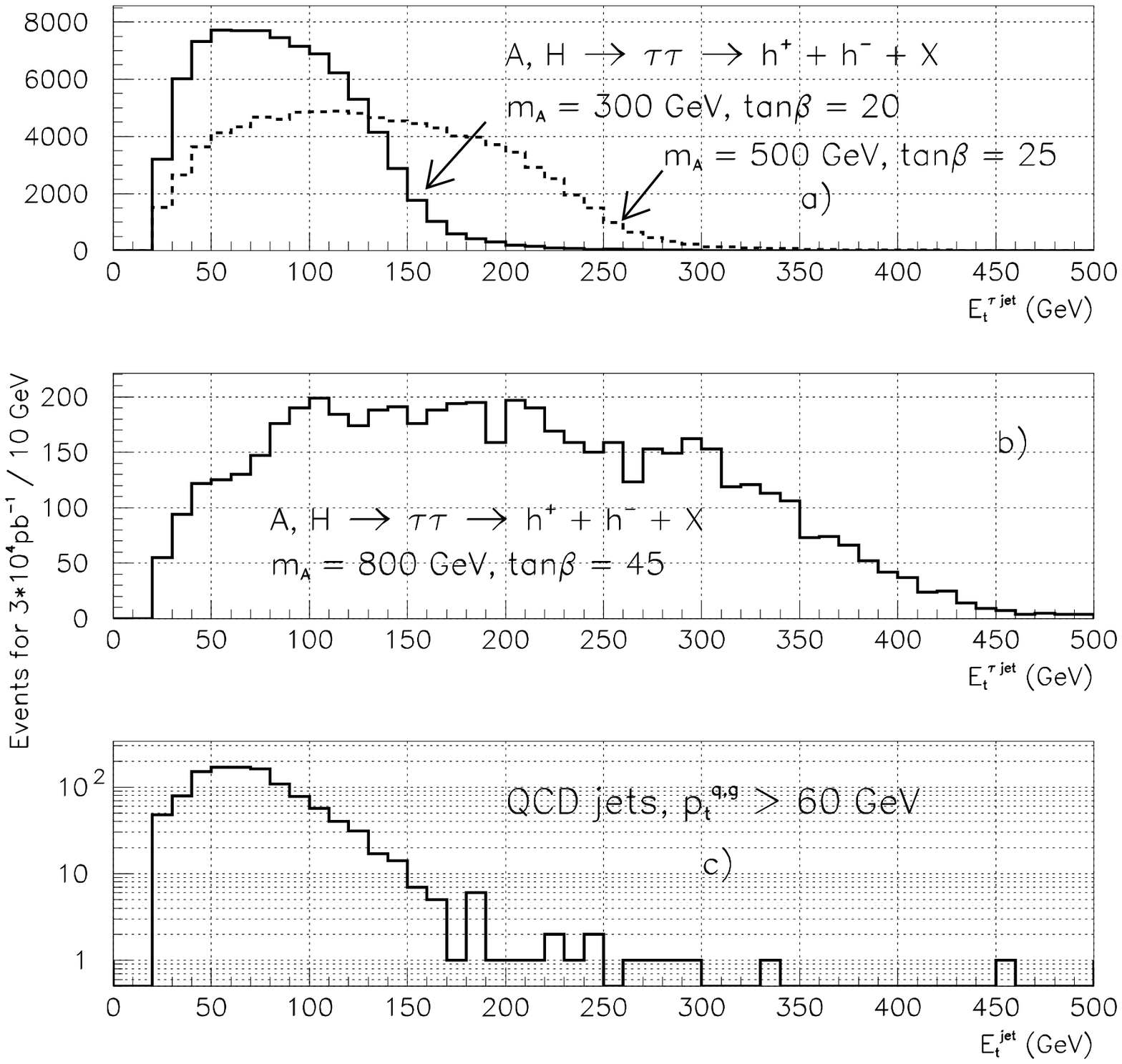}}
\caption
{a) Distribution of the visible transverse energy of the $\tau$ jet for $A,H$ \ra $~\tau\tau$ at $m_A$=300 GeV (solid histogram) and $m_A$=500 GeV (dashed histogram). 
b) The same for $A,H$ \ra $~\tau\tau$ at $m_A$=800 GeV.
c) Distribution of the transverse energy of the $\tau$ canditate (two hardest jets in the event) in the QCD jet events generated with $p_t^{q,g}>$60 GeV.}
\label{fig:1}
\end{figure}
\begin{figure}[htbp]
\centering
\resizebox{160mm}{210mm}
{\includegraphics{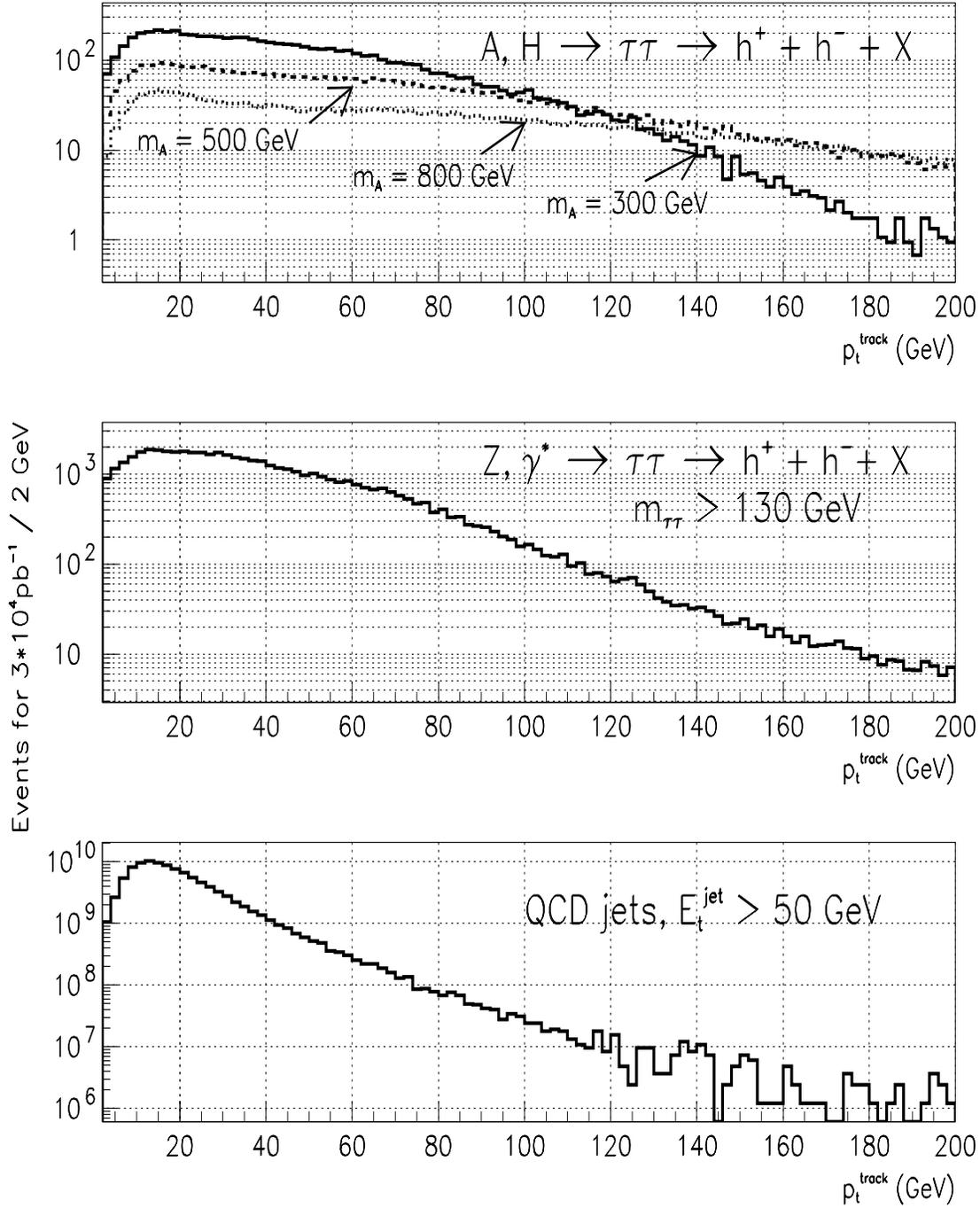}}
\caption
{a) Distribution of the transverse momentum of the isolated track in the jet with   $E_t >$ 60 GeV for $A,H$ \ra $~\tau\tau$ at $m_A$=300 GeV (solid histogram), $m_A$=500 GeV (dashed histogram) and $m_A$=800 GeV (dotted histogram). 
b) The same for $Z, \gamma^*$ \ra $~\tau\tau$ with $m_{\tau\tau}>$130 GeV,
c) The same for QCD 2-jet events}
\label{fig:2}
\end{figure}
\begin{figure}[htbp]
\centering
\resizebox{160mm}{210mm}
{\includegraphics{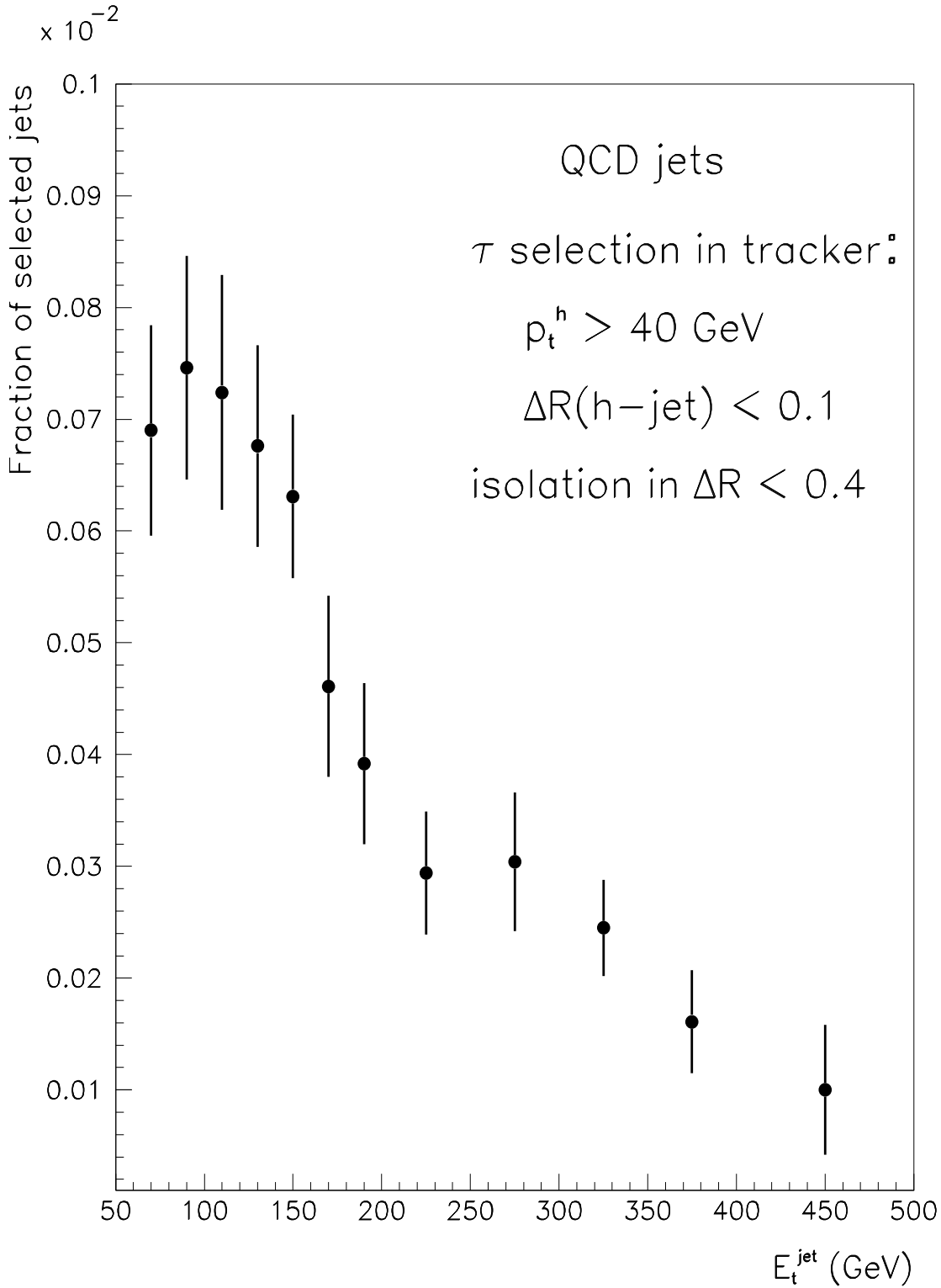}}
\caption
{Rejection factor for QCD jets as a function of $E_t^{jet}$. $\tau$ selection is performed in the tracker by requiring an isolated track with $p_t >$ 40 GeV within $\Delta R<$ 0.1 from the calorimeter jet axis. The isolation with a $p_t$ threshold of 1 GeV is extented to a larger cone of $\Delta R<$ 0.4.}
\label{fig:3}
\end{figure}
\begin{figure}[hbtp]
\centering
\resizebox{160mm}{210mm}
{\includegraphics{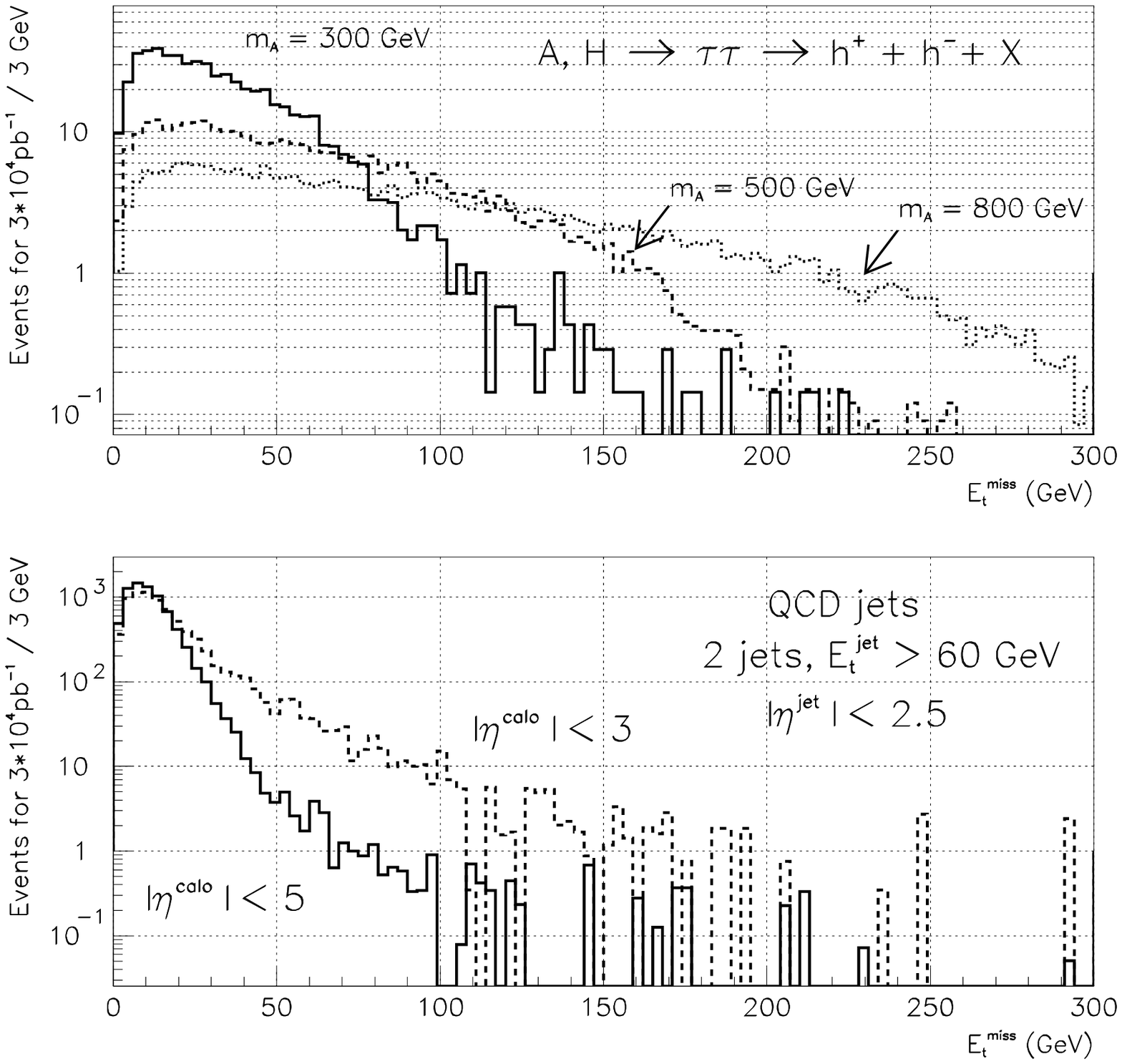}}
\caption
{a) Distribution of the missing transverse energy for $A,H$ \ra $~\tau\tau$ at $m_A$=300 GeV and $tan\beta$=15 (solid histogram), $m_A$=500 GeV and $tan\beta$=20 (dashed histogram) and $m_A$=800 GeV and $tan\beta$=45(dotted histogram). The detector response is simulated with CMSJET. No pile-up is included.  
b) The same for QCD jet events in the full $\eta$ range and for the central calorimeters only.}
    \label{fig:4}
\end{figure}
\begin{figure}[hbtp]
\centering
\resizebox{160mm}{210mm}
{\includegraphics{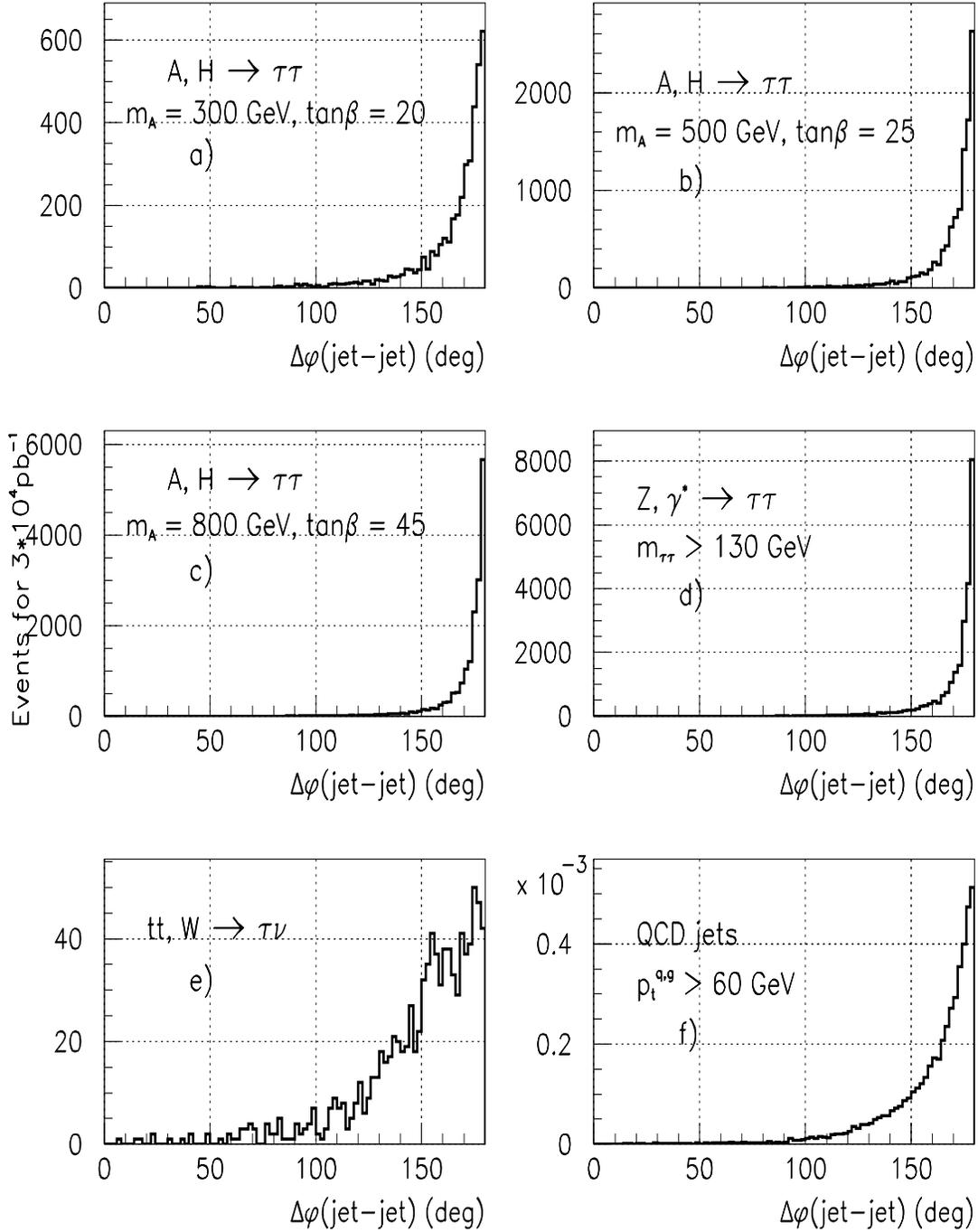}}
\caption
{Distribution of the $\Delta\phi$ angle in the transverse plane between the two $\tau$ jets for $A,H$ \ra $~\tau\tau$ at $m_A$=300 GeV and $tan\beta$=20 (a), $m_A$=500 GeV and $tan\beta$=25 (b), $m_A$=800 GeV and $tan\beta$=45 (c), $Z, \gamma^*$ \ra $~\tau\tau$ with $m_{\tau\tau}>$130 GeV (d), $t\overline{t}$ with $W$ \ra $~\tau\nu$ (e)  and QCD jet events (f).}
    \label{fig:5}
\end{figure}
\begin{figure}[hbtp]
\centering
\resizebox{160mm}{210mm}
{\includegraphics{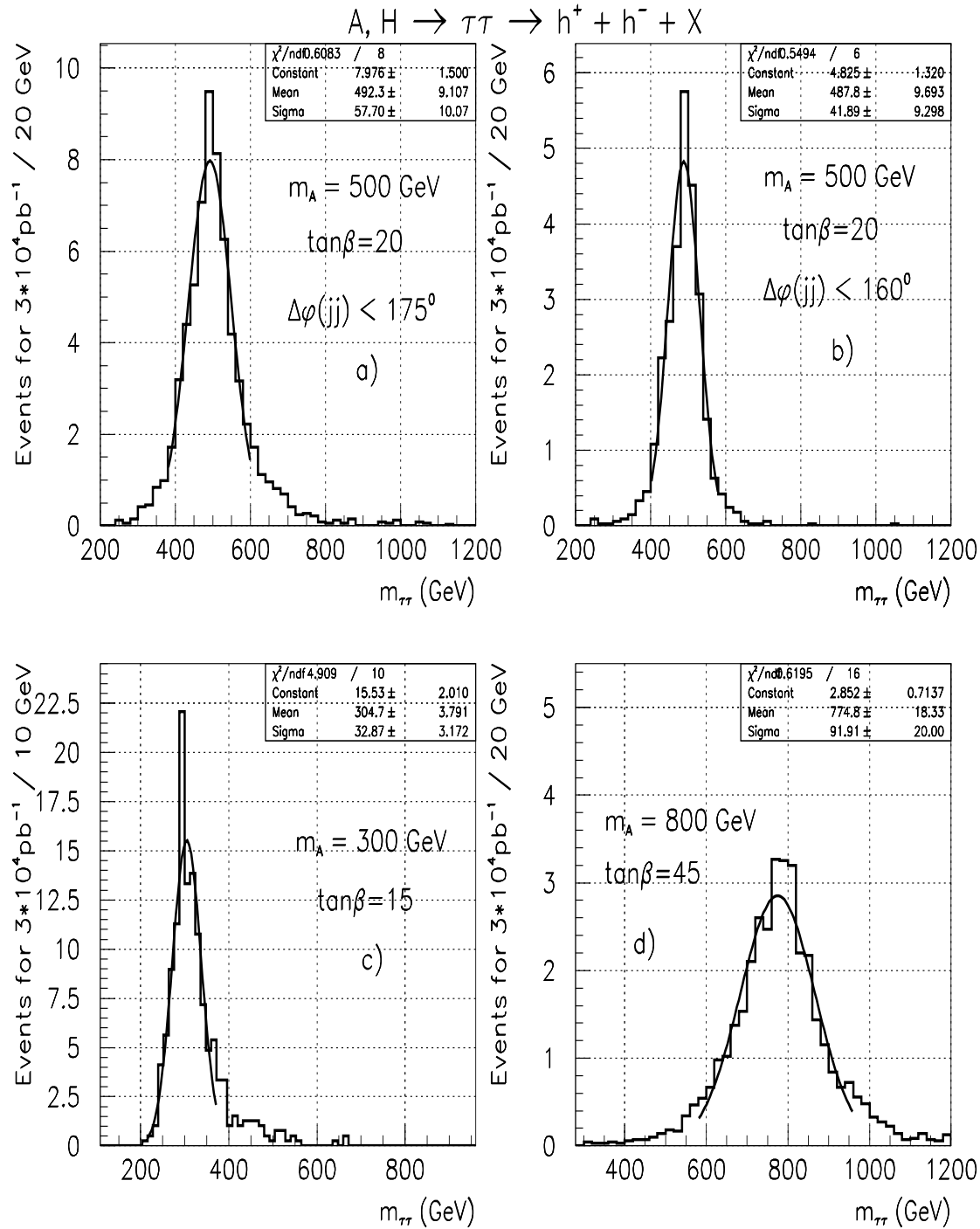}}
\caption
{a) Higgs mass reconstructed from the two $\tau$-jets and the $E_t^{miss}$ vector for $A,H$ \ra $~\tau\tau$ at $m_A$ = 500 GeV and $tan\beta$ = 20. The detector response is simulated with CMSJET. No pile-up is included.
b) The same as in a) but with $\Delta\phi(\tau1,\tau2)<$ 160$^0$.
c) Higgs mass reconstructed from the two $\tau$ jets and $E_t^{miss}$ vector for $A,H$ \ra $~\tau\tau$ at $m_A$ = 300 GeV and $tan\beta$ = 15.
d) Higgs mass reconstructed from the two $\tau$ jets and $E_t^{miss}$ vector for $A,H$ \ra $~\tau\tau$ at $m_A$ = 800 GeV and $tan\beta$ = 45.}
    \label{fig:6}
\end{figure}
\begin{figure}[hbtp]
\centering
\resizebox{160mm}{210mm}
{\includegraphics{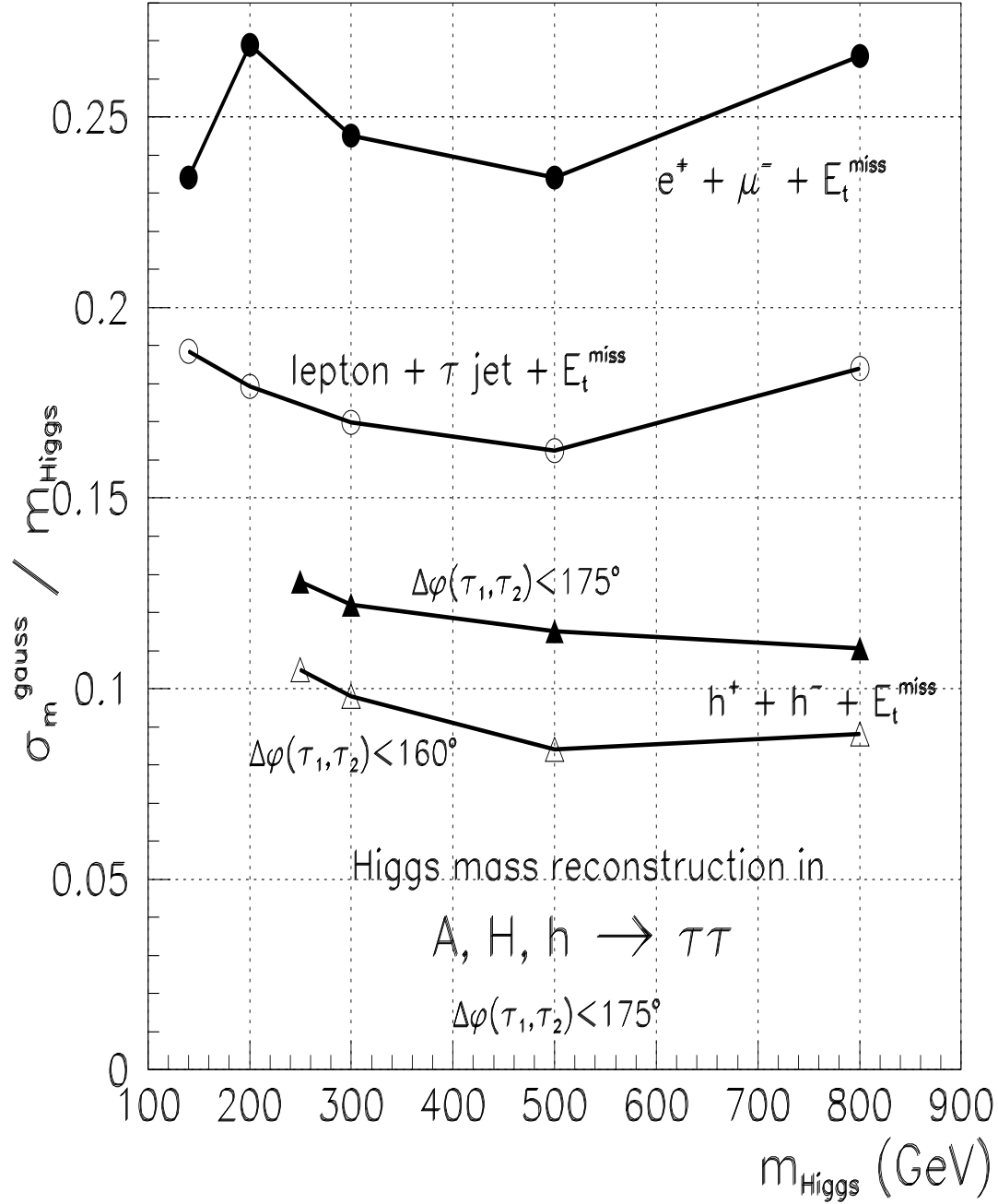}}
\caption
{relative mass resolution $\sigma_m/m$ of the reconstructed Higgs mass for the $A,H$ \ra $~\tau\tau$ in the $e^{\pm}$ +$\mu^{\mp}$, $lepton$ + $\tau$ jet and $h^{\pm}$ +$h^{\mp}$ final states as a function of Higgs mass. The event selection and mass reconstruction in the $e^{\pm}$ +$\mu^{\mp}$, $lepton$ + $\tau$ jet channels are discussed in refs. \cite{tauhlep}, \cite{tauemu}. The detector response is simulated with CMSJET. No pile-up is included.}
    \label{fig:7}
\end{figure}
\begin{figure}[hbtp]
\centering
\resizebox{160mm}{210mm}
{\includegraphics{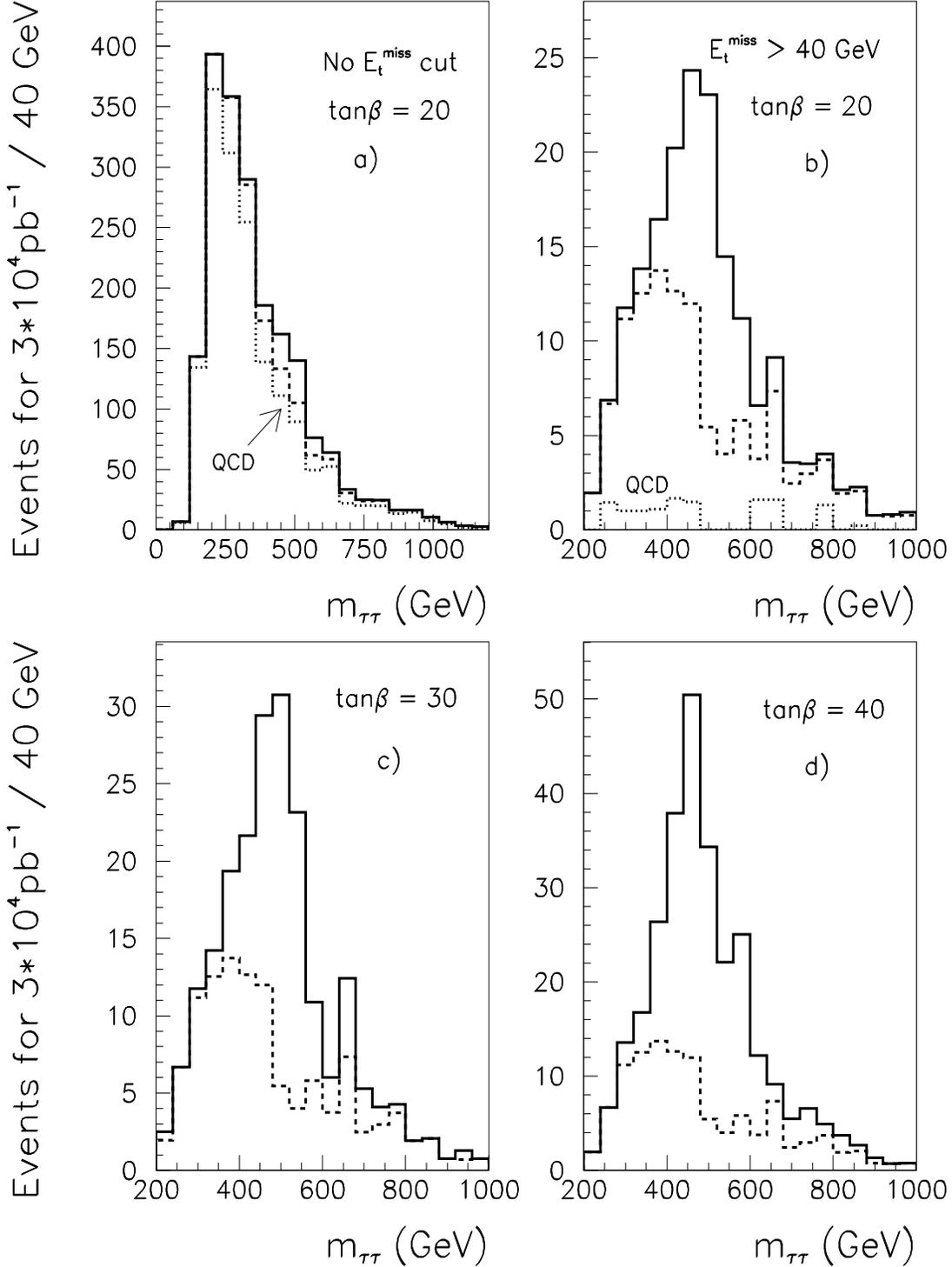}}
\caption
{a) Reconstructed Higgs mass for $A,H$ \ra $~\tau\tau$ at $m_A$ = 500 GeV and $tan\beta$ = 20 over the total background distribution for $3 \times 10^4pb^{-1}$. The statistical fluctuations correspond to the expected statistics for $3 \times 10^4pb^{-1}$. No cut is applied in $E_t^{miss}$.
b) The same as in a) but for $E_t^{miss}>$ 40 GeV.
c) Reconstructed Higgs mass for $A,H$ \ra $~\tau\tau$ at $m_A$ = 500 GeV and $tan\beta$ = 30 with the cut $E_t^{miss}>$ 40 GeV  over the total background distribution for $3 \times 10^4pb^{-1}$.
d) The same as in c) but for $tan\beta$ = 40.}
    \label{fig:8}
\end{figure}
\begin{figure}[hbtp]
\centering
\resizebox{160mm}{210mm}
{\includegraphics{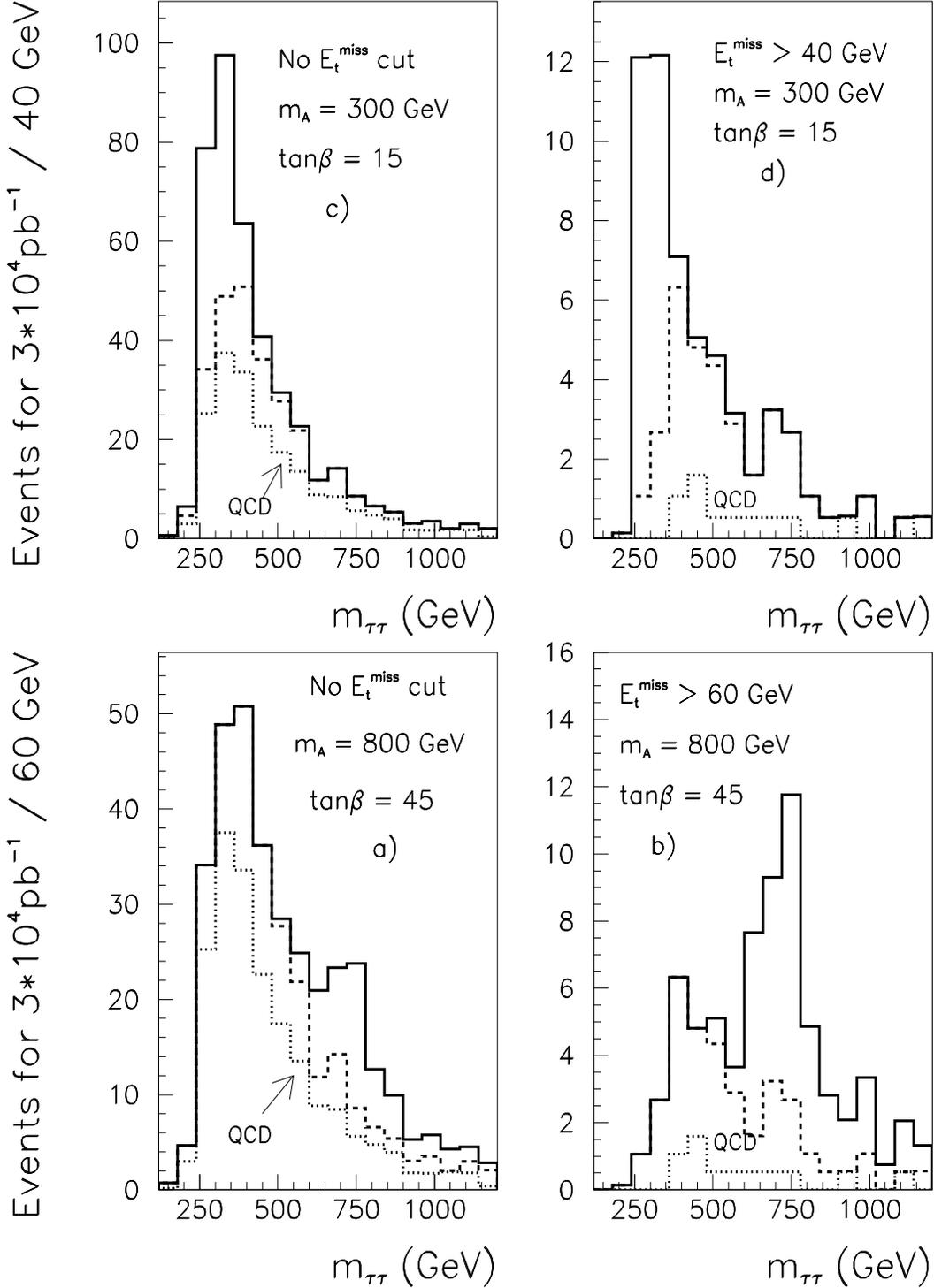}}
\caption
{a) Reconstructed Higgs mass for $A,H$ \ra $~\tau\tau$ at $m_A$ = 300 GeV and $tan\beta$ = 15 over the total background distribution for $3 \times 10^4pb^{-1}$.
b) The same as in a) but for $E_t^{miss}>$ 40 GeV.
c) Reconstructed Higgs mass for $A,H$ \ra $~\tau\tau$ at $m_A$ = 800 GeV and $tan\beta$ = 45 over the total background distribution for $3 \times 10^4pb^{-1}$. The statistical fluctuations correspond to the expected statistics for $3 \times 10^4pb^{-1}$. No cut is applied in $E_t^{miss}$.
d) The same as in c) but for $E_t^{miss}>$ 60 GeV.}
    \label{fig:9}
\end{figure}
\begin{figure}[hbtp]
\centering
\resizebox{160mm}{210mm}
{\includegraphics{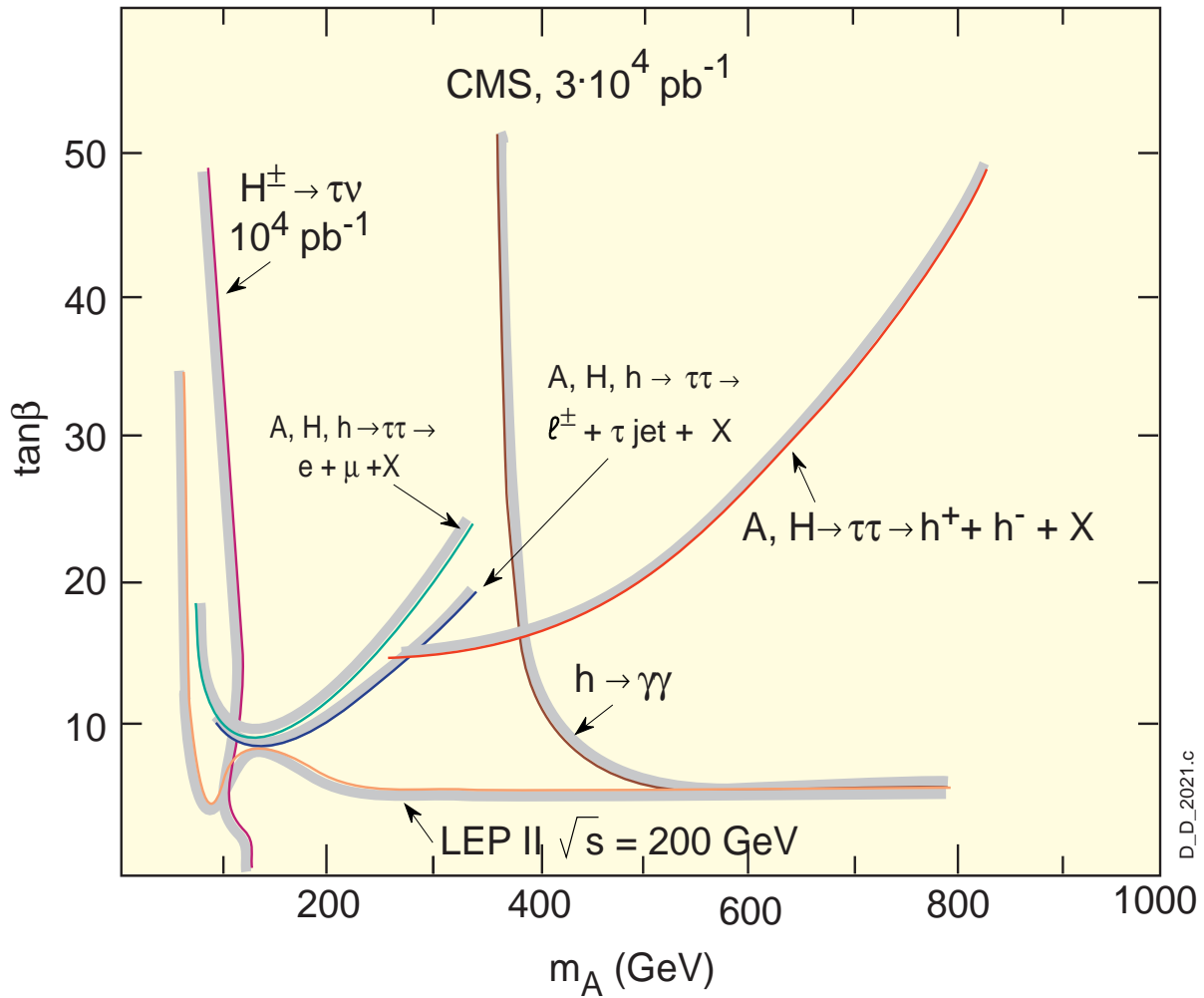}}
\caption
{Some 5$\sigma$-discovery limits for SUSY Higgses as a function of $m_A$ and $tan\beta$ for $3 \times 10^4pb^{-1}$ assuming no stop mixing.}
    \label{fig:10}
\end{figure}

\end{document}